\documentclass[a4paper,12pt]{article}
  \newtheorem{df}{Definition}[section]
  
  \newtheorem{lem}[df]{Lemma}
  
  \newtheorem{prop}[df]{Proposition}
  \newtheorem{cor}[df]{Corollary}
  \newtheorem{pbm}[df]{Problem}
  \font\hana=eufm10 scaled\magstep1
  \newfont{\fakefr}{eufm10}
  \def\sl{\mbox{{\hana sl}}}
  \def\su{\mbox{{\hana su}}}

  \newcommand{\adag}{a^{\dag{}}}
  \newcommand{\Adag}{A^{\dag{}}}
  \newcommand{\ra}{\rangle}
  \newcommand{\la}{\langle}
  \makeatletter
   
   \@addtoreset{equation}{section}
  \makeatother

\title{A Universal Disentangling Formula for \\
Coherent States of Perelomov's Type}
\author{Kazuyuki FUJII\thanks{Department of Mathematics, 
  Yokohama City University, 
  Yokohama 236, 
  Japan, \endgraf 
  {\it E-mail address}: fujii{\char'100}yokohama-cu.ac.jp} \ and 
  Tatsuo SUZUKI\thanks{Department of Mathematics, 
  Waseda University, 
  Tokyo 169, 
  Japan, \endgraf 
  {\it E-mail address}: suzukita{\char'100}mse.waseda.ac.jp}}
\date{}
\begin{document}
\maketitle
\begin{abstract}
A universal disentangling formula (such as the Baker-Campbell-Hausdorff one) for 
coherent states of Perelomov's type ($ |z \ra =
\exp (z\Adag -\bar{z}A) |0 \ra $) 
which are defined for generalized oscillator algebras is given.
\end{abstract}
%%%%%%%%%%%%%%%%%%%%%%%%%%%%%%%%%%%%%%%%%%%%%%%%%%%%%%%%%%%%%%%%
 \section{Introduction}
Coherent states play an important role in quantum physics, in particular 
quantum optics, see \cite{KS} and its references. They are very useful in performing 
stationary-phase approximations to path integral, see \cite{KS} or \cite{FKSF1}, \cite{FKSF2}, \cite{FKS}. 

To begin with, let $a$ and $\adag$ be canonical annihilation and 
creation operators in 
the harmonic oscillator. We set $|0\ra $ a normalized vacuum ($a|0 \ra =0$ and 
$\la 0|0 \ra =1$). The (normalized) coherent state $|z \ra $ is defined as an 
eigenfunction of $a$:
\begin{equation}
a|z \ra =z|z \ra  \quad \mbox{for} \quad z \in \mathbf{C}.
\end{equation}
This $|z \ra $ is written as 
$$
 |z \ra = \mbox{e}^{-\frac{|z|^2}{2}}\mbox{e}^{z\adag}|0 \ra 
        = \mbox{e}^{z\adag-\bar{z}a}|0 \ra .
$$
In obtaining the last equality, we have used the fundamental 
Baker-Campbell-Hausdorff formula
\begin{equation}
 \mbox{e}^{A+B}=\mbox{e}^{-\frac12[A,B]}\mbox{e}^{A}\mbox{e}^{B}
  \label{eqn:0-2}
\end{equation}
whenever $[A,[A,B]]=[B,[A,B]]=0$. 
Maybe this is the first disentangling formula. 

Next, we denote by $\{ K_+, K_-, K_3 \}$ operators expressing spin $K$ 
($K \geq 1/2$) representation of $\su (1,1)$. Then Perelomov defined 
the following coherent state
\begin{equation}
 |z \ra =\mbox{e}^{zK_+ -\bar{z}K_-} |K,0 \ra  
  \quad \mbox{for} \quad z \in \mathbf{C},
\end{equation}
see \cite{Pe}, \cite{KS}. To this system, we cannot apply 
the Baker-Campbell-Hausdorff formula (\ref{eqn:0-2}), 
but have a disentangling formula 
of another type \cite{Pe}, \cite{HS}: For $\zeta =(z \tanh |z|)/|z|$ \ 
($z \in \mathbf{C}$)
\begin{equation}
 \mbox{e}^{zK_+ -\bar{z}K_-} = 
  \mbox{e}^{\zeta K_+}\mbox{e}^{\log (1-|\zeta|^2)K_3}
  \mbox{e}^{-\bar{\zeta}K_-}.
\end{equation}
{}From this we have 
\begin{equation}
 |z \ra =(1-|\zeta|^2)^K \mbox{e}^{\zeta K_+}|K,0 \ra . 
\end{equation}

In view of these two cases, a following question arises. For a generalized oscillator 
algebra $\mathcal{A}=\{ 1,A,\Adag,N \}$, we can consider a coherent state 
of Perelomov's type 
\begin{equation}
 |z \ra =\mbox{e}^{z\Adag -\bar{z}A} |0 \ra  
  \quad \mbox{for} \quad z \in \mathbf{C}.
\end{equation}
What kind of disentangling formulas apply to this system? Such a formula 
is not known as far as we know. 

In this letter, we give a disentangling formula to this system and recover 
the two results stated above as particular case. Our formula is expressed as 
an infinite series, so it is not easy to find a function form such as 
$\mbox{e}^{-|z|^2},\mbox{tanh}|z|$, etc, except for 
some simple cases. Therefore our work is a first step to obtain the real 
universal disentangling formula. Further study will be expected. The details 
and further developments of this letter will be published elsewhere, \cite{FS}. 
%%%%%%%%%%%%%%%%%%%%%%%%%%%%%%%%%%%%%%%%%%%%%%%%%%%%%%%%%%%%%%
 \section{Coherent States}\label{section:1}
Let us make a short review of coherent states, \cite{KS}. We set 
$\{ 1,a,\adag, N \equiv \adag a \}$ the oscillator algebra, 
\begin{equation}
 [N,\adag]=\adag, \ [N,a]=-a, \ [a,\adag]=1.
\end{equation}
The Fock space on which $a$ and $\adag$ act is $\mathcal{H} \equiv 
\{ |n \ra  \ |n \geq 0 \}$ and whose actions are
\begin{eqnarray}
 \adag |n \ra  &=& \sqrt{n+1}|n+1 \ra , \nonumber\\
 a \ |n \ra      &=& \sqrt{n-1}|n-1 \ra , \label{eqn:1-2}\\
 N|n \ra       &=& n|n \ra , \nonumber
\end{eqnarray}
where $|0 \ra $ is a normalized vacuum ($a|0 \ra =0$ and $ \la 0|0 \ra =1$).
 {}From (\ref{eqn:1-2}), states $|n \ra $ are given by 
\begin{equation}
 |n \ra =\frac{(\adag)^n}{\sqrt{n!}}|0 \ra .
  \label{eqn:1-3}
\end{equation}
These states satisfy the orthogonality and completeness conditions 
\begin{equation}
  \la m|n \ra =\delta_{mn}, \quad \sum_{n=0}^{\infty}|n \ra  \la n|=1.
\end{equation}
An unnormalized coherent state $|z)$ is defined as 
\begin{equation}
 a|z)=z|z) \quad \mbox{for} \quad z \in \mathbf{C}.
  \label{eqn:1-5}
\end{equation}
This equation is easily solved to be 
\begin{equation}
 |z)=\sum_{n=0}^{\infty} \frac{z^n}{\sqrt{n!}}|n \ra ,
\end{equation}
or from (\ref{eqn:1-3})
\begin{equation}
 |z)=\sum_{n=0}^{\infty} \frac{(z\adag)^n}{n!}|0 \ra 
        =\mbox{e}^{z\adag}|0 \ra .
\end{equation}
 {}From $(z|z)=\mbox{e}^{|z|^2}$, the normalized coherent state 
becomes 
\begin{equation}
 |z \ra  \equiv \frac{1}{\sqrt{(z|z)}} \mbox{e}^{z\adag}|0 \ra 
      =\mbox{e}^{-\frac{|z|^2}{2}} \mbox{e}^{z\adag}|0 \ra . 
  \label{eqn:1-8}
\end{equation}
Therefore $|z \ra $ is written as 
\begin{equation}
 |z \ra =\mbox{e}^{z\adag -\bar{z}a} |0 \ra . 
   \label{eqn:1-9}
\end{equation}
Here we have used the fundamental Baker-Campbell-Hausdorff formula
\begin{equation}
 \mbox{e}^{A+B}=\mbox{e}^{-\frac12[A,B]}\mbox{e}^{A}\mbox{e}^{B}
  \label{eqn:1-10}
\end{equation}
whenever $ [A,[A,B]]=[B,[A,B]]=0$. 
(\ref{eqn:1-10}) is a first disentangling formula (it is not easy to derive 
(\ref{eqn:1-9}) from (\ref{eqn:1-8}) without knowing this formula). 
In what follows we call coherent states 
(\ref{eqn:1-9}) a Perelomov's type  
and study in detail. On the other hand, we call (\ref{eqn:1-5}) a 
Barut-Girardello's one, but we are not concerned with it in this letter, 
see \cite{BG}, \cite{FF1}, \cite{FF2}. 
%%%%%%%%%%%%%%%%%%%%%%%%%%%%%%%%%%%%%%%%%%%%%%%%%%%%%%%
 \section{The Disentangling Formula for Perelomov's Coherent States}
  \label{section:2}
Let $\{ k_+, k_-, k_3 \}$ be a Weyl basis of Lie algebra $\su (1,1) 
\subset \sl (2,\mathbf{C})$, 
$$
 k_+ = \left(
        \begin{array}{cc}
               0 & 1 \\
               0 & 0 \\
        \end{array}
       \right),
 \quad 
 k_- = \left(
        \begin{array}{cc}
               0 & 0 \\
              -1 & 0 \\
        \end{array}
       \right),
 \quad 
 k_3 = \frac12 
       \left(
        \begin{array}{cc}
               1 &  0  \\
               0 & -1 \\
        \end{array}
       \right). 
$$
Then we have 
\begin{equation}
 [k_3, k_+]=k_+, \quad [k_3, k_-]=-k_-, \quad [k_-, k_+]=2k_3.
\end{equation}
Next we consider a spin $K$ ($\geq 1/2$) representation of $\su (1,1) 
\subset \sl (2,\mathbf{C})$ and set its generators 
$\{ K_+, K_-, K_3 \}$, 
\begin{equation}
 [K_3, K_+]=K_+, \quad [K_3, K_-]=-K_-, \quad [K_-, K_+]=2K_3.
\end{equation}
We note that this (unitary) representation is infinite dimensional. 
The Fock space on which $\{ K_+, K_-, K_3 \}$ act is 
$\mathcal{H}_K \equiv 
\{ |K,n \ra  \ |n \geq 0 \}$ and whose actions are
\begin{eqnarray}
 K_+ |K,n \ra  &=& \sqrt{(n+1)(2K+n)}|K,n+1 \ra , \nonumber\\
 K_- |K,n \ra  &=& \sqrt{n(2K+n-1)}|K,n-1 \ra , \label{eqn:2-3}\\
 K_3 |K,n \ra  &=& (K+n)|K,n \ra , \nonumber
\end{eqnarray}
where $|K,0 \ra $ is a normalized vacuum ($K_-|K,0 \ra =0$ and 
$ \la K,0|K,0 \ra =1$). We have written $|K,0 \ra $ instead of $|0 \ra $ 
to emphasize 
the spin $K$ representation, see \cite{FKSF1}. {}From (\ref{eqn:2-3}), states 
$|K,n \ra $ are given by 
\begin{equation}
 |K,n \ra =\frac{(K_+)^n}{\sqrt{n!(2K)_n}}|K,0 \ra ,
  \label{eqn:2-4}
\end{equation}
where $(a)_n$ is the Pochammer's notation
$$
 (a)_n \equiv  a(a+1) \cdots (a+n-1).
$$
These states satisfy the orthogonality and completeness conditions 
\begin{equation}
  \la K,m|K,n \ra =\delta_{mn}, 
 \quad \sum_{n=0}^{\infty}|K,n \ra  \la K,n|=1_K.
\end{equation}
Now let us consider the coherent state of Perelomov's type 
\begin{equation}
 |z \ra  \equiv \mbox{e}^{zK_+ -\bar{z}K_-} |K,0 \ra  
  \quad \mbox{for} \quad z \in \mathbf{C}.
   \label{eqn:2-6}
\end{equation}
This is an extension of (\ref{eqn:1-9}). We want to disentangle 
(\ref{eqn:2-6}). But, since 
$$
[K_+,[K_+,K_-]]=2K_+ \ne 0, \quad [K_-,[K_+,K_-]]=-2K_- \ne 0, 
$$
we cannot use the Baker-Campbell-Hausdorff formula 
(\ref{eqn:1-10}). In spite of this fact, we fortunately have an another 
disentangling formula, \cite{Pe}, \cite{HS}:
\begin{lem}
For $\zeta =(z \tanh |z|)/|z|$ \ 
$(z \in \mathbf{C})$ 
\begin{equation}
 \mbox{e}^{zK_+ -\bar{z}K_-} = 
  \mbox{e}^{\zeta K_+}\mbox{e}^{\log (1-|\zeta|^2)K_3}
  \mbox{e}^{-\bar{\zeta}K_-}.
 \label{eqn:2-7}
\end{equation}
\end{lem}
Therefore we obtain from (\ref{eqn:2-3})
\begin{equation}
 |z \ra =(1-|\zeta|^2)^K \mbox{e}^{\zeta K_+}|K,0 \ra . 
  \label{eqn:2-8}
\end{equation}
A comment here is in order. In this lemma, $\zeta$ plays 
an essential role and, moreover, $\zeta$ is an element of Poincare disk;
$$
 \zeta \in \mathbf{D} \equiv 
  \{ z \in \mathbf{C}|\ |z|<1 \}
$$
in contrast to $z \in \mathbf{C}$. 
This difference is important, see \cite{Pe}. 

As a consequence of section \ref{section:1} and \ref{section:2}, 
coherent states of Perelomov's type have different disentangling formulas 
according to algebras. 

Now we come to a natural question: Is there a universal disentangling formula 
which does not depend on characteristics of algebras? In the next section, 
we treat this question and give some answer. 
%%%%%%%%%%%%%%%%%%%%%%%%%%%%%%%%%%%%%%%%%%%%%%%%%%%%%%%%
 \section{A Universal Disentangling Formula}\label{section:3}
Let $\mathcal{A} \equiv \{ 1,A,\Adag,N \}$ be a generalized oscillator 
algebra \cite{QV}, \cite{KP},
\begin{equation}
 [N,\Adag]=\Adag, \quad [N,A]=-A, \quad [A,\Adag]=F(N+1)-F(N),
\end{equation}
where $F$ is an entire function satisfying $F(0)=0$ and $F(n)>0$. 
The Fock space on which $A$ and $\Adag$ act is $\{ |n \ra \ |n\geq 0 \}$
and whose actions are
\begin{eqnarray}
 \Adag |n \ra  &=& \sqrt{F(n+1)}|n+1 \ra , \nonumber\\
 A     |n \ra  &=& \sqrt{F(n)}|n-1 \ra , \label{eqn:3-2}\\
 N     |n \ra  &=& n|n \ra , \nonumber
\end{eqnarray}
where $|0 \ra $ is a normalized vacuum. {}From (\ref{eqn:3-2}), states 
$|n \ra $ are given by 
\begin{equation}
 |n \ra  = \frac{(\Adag)^n}{\sqrt{\prod_{j=1}^{n}F(j)}}|0 \ra . 
\end{equation}
These states satisfy the orthogonality and completeness conditions 
\begin{equation}
  \la m|n \ra =\delta_{mn}, \quad \sum_{n=0}^{\infty}|n \ra  \la n|=1.
\end{equation}
Now we define a coherent state of Perelomov's type as 
\begin{equation}
 |z \ra  \equiv \mbox{e}^{z\Adag-\bar{z}A}|0 \ra  
  \quad \mbox{for} \quad z \in \mathbf{C}.
   \label{eqn:3-5}
\end{equation}
Though we want to disentangle (\ref{eqn:3-5}), useful formula such as 
(\ref{eqn:1-10}) in section \ref{section:1} or (\ref{eqn:2-7}) in section 
\ref{section:2} are not known for a general algebra $\mathcal{A}$ 
as far as we know. Therefore we must construct a disentangling formula 
for (\ref{eqn:3-5}). 

Let us describe our main result. 
\begin{prop}\label{prop:3-1}
We have 
\begin{equation}
 |z \ra  =
   \sum_{n=0}^{\infty}
    \left\{
     \sum_{j=0}^{\infty}\frac{n!}{(n+2j)!}
      \Delta (n+1,j)(-|z|^2)^j 
    \right\} \frac{(z\Adag)^n}{n!}|0 \ra ,
 \label{eqn:3-6}
\end{equation}
where $\Delta (n+1,j)$ is defined as 
\begin{eqnarray}
 && \Delta (n+1,0)=1, \nonumber\\
 && \Delta (n+1,j)=\sum_{k_1=1}^{n+1}F(k_1)\sum_{k_2=1}^{k_1+1}F(k_2)
               \cdots \sum_{k_j=1}^{k_{j-1}+1}F(k_j).
 \label{eqn:3-7}
\end{eqnarray}
\end{prop}
This is just what we call a universal disentangling formula in the title 
of this letter. We note that $\Delta (n+1,j)$ satisfy the recurrence formula 
\begin{equation}
 \Delta (n+1,j)=\Delta (n,j)+F(n+1)\Delta (n+2,j-1).
\end{equation}
It is interesting that (\ref{eqn:3-7}) is a bit similar to a discrete form 
of path-ordered integral. 

To start with, we treat the harmonic oscillator $F(n)=n$. 
In this case we show that (\ref{eqn:3-6}) coincides with 
(\ref{eqn:1-8}). Applying the well-known formula
$$
\sum_{k=1}^{n}(k)_j = \frac{(n)_{j+1}}{j+1},
$$
where $(k)_j$ is the Pochammer notation, to (\ref{eqn:3-7}) repeatedly, 
we have 
\begin{equation}
 \Delta (n+1,j)=\frac{(n+1)_{2j}}{(2j)!!}=\frac{(n+2j)!}{n!2^j j!},
\end{equation}
so that 
\begin{equation}
 \sum_{j=0}^{\infty}\frac{n!}{(n+2j)!}\Delta (n+1,j)(-|z|^2)^j 
  = \sum_{j=0}^{\infty}\frac{(-|z|^2)^j}{2^j j!} 
  = \mbox{e}^{-\frac{|z|^2}{2}}. 
\end{equation}
Because this quantity is independent of $n$, we obtain 
the following corollary.
\begin{cor}
$$
 |z \ra =\mbox{e}^{-\frac{|z|^2}{2}}\mbox{e}^{z\adag}|0 \ra . 
$$
\end{cor}
We could recover the result in section \ref{section:1}. 

Next, we consider the spin $K$ representations of $\su (1,1)$. 
We show that (\ref{eqn:3-6}) coincides with (\ref{eqn:2-8}). In this case 
\begin{equation}
 F(n)=n(2K+n-1),
  \label{eqn:3-11}
\end{equation}
we have only to calculate $\Delta (n+1,j)$. But 
\begin{eqnarray*}
 && \Delta (n+1,0)=1, \\
 && \Delta (n+1,1)=\sum_{k=1}^{n+1} k(2K+k-1)
            =\frac{(n+1)(n+2)(n+3K)}{3}. 
\end{eqnarray*}
The situation becomes difficult more and more for $j \geq 2$. 
\begin{pbm}
Is it possible to determine $\Delta (n+1,j)$ completely?
\end{pbm}
Here we change our strategy. In the following, we shall develop 
a general theory. 
\begin{eqnarray}
 && \mbox{e}^{z\Adag-\bar{z}A}|0 \ra  \nonumber\\
 && = \sum_{n=0}^{\infty}
    \left\{
     \sum_{j=0}^{\infty}(-1)^j \frac{n!}{(n+2j)!}
      \Delta (n+1,j) |z|^{2j} 
    \right\} \frac{(z\Adag)^n}{n!}|0 \ra  \nonumber\\
 && = \sum_{n=0}^{\infty}
    \left\{
     \sum_{j=0}^{\infty}(-1)^j \frac{n!}{(n+2j)!}
      \Delta (n+1,j) |z|^{2j+n} 
    \right\} \frac{1}{n!}
    \left(
     \frac{z}{|z|}\Adag 
    \right)^n |0 \ra  \nonumber\\
 && \equiv \sum_{n=0}^{\infty} I_n(|z|)\ \frac{1}{n!}
    \left(
     \frac{z}{|z|}\Adag 
    \right)^n |0 \ra .
 \label{eqn:3-12}
\end{eqnarray}
Setting $r=|z|$ for simplicity, 
\begin{equation}
 I_n(r) = \sum_{j=0}^{\infty}(-1)^j \frac{n!}{(n+2j)!}
           \Delta (n+1,j) r^{2j+n}. 
 \label{eqn:3-13}
\end{equation}
Here let us determine a differential equation which 
$\{ I_n(r) \ |n\geq 0 \}$ satisfy instead of determining themselves. 
\begin{prop}
For $n \geq 0$, 
\begin{equation}
 \frac{d}{dr}I_n(r)=
  nI_{n-1}(r)-\frac{F(n+1)}{n+1}I_{n+1}(r)
   \label{eqn:3-14}
\end{equation}
and 
\begin{equation}
 I_n(r) \sim r^n \quad \mbox{for} \quad 0<r \ll 1.
  \label{eqn:3-15}
\end{equation}
\end{prop}
 {}From this proposition, if we can solve (\ref{eqn:3-14}) with boundary 
conditions (\ref{eqn:3-15}), then we can determine $|z \ra $ making use of 
(\ref{eqn:3-12}). 

Now we go back to the case (\ref{eqn:3-11}); $F(n)=n(2K+n-1)$. In this case 
(\ref{eqn:3-14}) becomes 
\begin{equation}
 \frac{d}{dr}I_n = nI_{n-1}-(2K+n)I_{n+1}. 
\end{equation}
Then the solution with (\ref{eqn:3-15}) are known to be 
\begin{eqnarray}
 I_n(r) &=& (\cosh r)^{-2K-n}(\sinh r)^n \nonumber\\
        &=& (\cosh r)^{-2K}(\tanh r)^n 
         =  (1-\tanh^2 r)^K (\tanh r)^n. 
  \label{eqn:3-17}
\end{eqnarray}
Therefore by (\ref{eqn:3-12}) 
\begin{eqnarray}
 |z \ra  &=& 
  (1-\tanh^2 |z|)^K 
   \sum_{n=0}^{\infty}(\tanh |z|)^n \frac{1}{n!}
    \left(
     \frac{z}{|z|}K_+
    \right)^n |K,0 \ra  \nonumber\\
 &=& (1-\tanh^2 |z|)^K 
      \mbox{e}^{\tanh |z| \frac{z}{|z|} K_+}|K,0 \ra .
\end{eqnarray}
That is, setting $\zeta = (z \tanh |z|)/|z|$, 
\begin{cor}
 we have 
$$
 |z \ra =(1-|\zeta|^2)^K \mbox{e}^{\zeta K_+}|K,0 \ra . 
$$
\end{cor}
We could also recover the result in section \ref{section:2}. 
%%%%%%%%%%%%%%%%%%%%%%%%%%%%%%%%%%%%%%%%%%%%%%%%%%%
 \section{An Application to Elementary Number Theory}
In this section, we apply the result of section \ref{section:3} to 
elementary number theory. For the case $F(n)=n(2K+n-1)$, we gave two 
different disentangling formulas to the coherent state $|z \ra $; 
(\ref{eqn:2-8}) and Proposition \ref{prop:3-1}. We in particular consider 
the case $n=0$ in (\ref{eqn:3-13}): 
\begin{eqnarray}
 && I_0(r) = \sum_{j=0}^{\infty}(-1)^j \Delta (1,j) \frac{r^{2j}}{(2j)!}
     \qquad \mbox{by (\ref{eqn:3-13})}, \\
 && I_0(r)=(\cosh r)^{-2K}
     \qquad \qquad \qquad \mbox{by (\ref{eqn:3-17})}.
\end{eqnarray}
 {}From these we have a simple formula 
\begin{equation}
 (\cosh r)^{-2K}
  = \sum_{j=0}^{\infty}(-1)^j \Delta (1,j) \frac{r^{2j}}{(2j)!}, 
 \label{eqn:4-3}
\end{equation}
where 
\begin{eqnarray}
 && \Delta (n,j)=\sum_{k=1}^n F(k)\Delta (k+1,j-1), \\
 && F(k)=k(2K+k-1), \quad K \geq \frac12 . 
\end{eqnarray}
We moreover set $K=1/2$. 
\begin{equation}
 \frac{1}{\cosh r}
  = \sum_{j=0}^{\infty}(-1)^j \Delta (1,j) \frac{r^{2j}}{(2j)!}, 
 \label{eqn:4-6}
\end{equation}
\begin{equation}
 F(k)=k^2.
\end{equation}
In this case, let us calculate $\Delta (1,1) \sim \Delta (1,3)$ explicitly. 
\begin{eqnarray}
 && \Delta (1,1)=F(1)=1, \nonumber\\
 && \Delta (1,2)=F(1)(F(1)+F(2))=1+4=5, \nonumber\\
 && \Delta (1,3)
      =F(1)^2(F(1)+F(2))+F(1)F(2)(F(1)+F(2)+F(3)) \nonumber\\
  && \qquad \quad 
      =5+4 \times 14=61. 
 \label{eqn:4-8}
\end{eqnarray}
On the other hand, the expression of $1/\cosh r$ is well-known to be 
\begin{equation}
 \frac{1}{\cosh r}
  = \sum_{j=0}^{\infty}(-1)^j E_j \frac{r^{2j}}{(2j)!}, 
 \label{eqn:4-9}
\end{equation}
where $E_j$ are the Euler numbers and are given by the recurrence formula 
\begin{equation}
 E_0=1, \quad \sum_{k=0}^n (-1)^k 
               \left(
                \begin{array}{c}
                  2n \\
                  2k \\
                \end{array}
               \right) E_{n-k}=0. 
 \label{eqn:4-10}
\end{equation}
For example, we have 
\begin{eqnarray}
 && E_1=E_0=1, \nonumber\\
 && E_2=\left( \begin{array}{c} 4 \\ 2 \\ \end{array} \right) E_1-E_0
       =6-1=5, \nonumber\\
 && E_3=\left( \begin{array}{c} 6 \\ 4 \\ \end{array} \right) E_2-
        \left( \begin{array}{c} 6 \\ 2 \\ \end{array} \right) E_1+E_0
         \nonumber\\
  && \quad =15 \times 5-15 \times 1+1=61.
 \label{eqn:4-11}
\end{eqnarray}
Comparing (\ref{eqn:4-6}) with (\ref{eqn:4-9}), 
\begin{lem}
we have 
\begin{equation}
 \Delta (1,j)|_{K=1/2}=E_j. 
\end{equation}
\end{lem}
But our $\Delta (1,j)$ and $E_j$ are clearly distinct expressions (compare 
(\ref{eqn:4-8}) with (\ref{eqn:4-11})). In fact, it is not easy to prove 
that $\{ \Delta (1,j)|_{K=1/2} \}$ satisfy the equation (\ref{eqn:4-10}) 
directly (Try a direct proof). 
Moreover, if we define generalized Euler numbers $\tilde{E}_j$ by 
\begin{equation}
 \frac{1}{(\cosh r)^{2K}}
  = \sum_{j=0}^{\infty}(-1)^j \tilde{E}_j \frac{r^{2j}}{(2j)!}, 
\end{equation}
then $\Delta (1,j)$ in (\ref{eqn:4-3}) is an another expression of  
$\tilde{E}_j$. 

In conclusion, our work could give an another expression to some facts in 
elementary number theory from a viewpoint of representation theory. We can 
expect more applications. 
%%%%%%%%%%%%%%%%%%%%%%%%%%%%%%%%%%%%%%%%%%%%%%%%%%%%%%%%%%%%
 \section{Discussion}
In this letter, we defined coherent states of Perelomov's type for a 
generalized oscillator algebra, and gave a disentangling formula to them. 
Since our formula is given by complicated infinite series, it is not easy 
to find a function form except for some simple cases. But our formula doesn't 
depend on special skills of algebras, so it is universal ! 

We have not yet applied our formula to nonlinear algebras defined in \cite{CJT}, \cite{JR} 
(in which coherent states of Barut-Girardello's type have been constructed). 
This is an interesting subject. 

Our work is nothing but a first step to obtain 
the real universal disentangling formula 
for coherent states of Perelomov's type. We hope that based on our 
work the ultimate disentangling formula will be obtained in the near future. 
%%%%%%%%%%%%%%%%%%%%%%%%%%%%%%%%%%%%%%%%%%%%%%%%%%%
 \section*{Acknowledgements}
Kazuyuki Fujii would like to thank Prof. Georg Junker for sending their 
preprints to us. Their works are one of our motivation to extend the 
disentangling formula. Tatsuo Suzuki is very grateful to Saburo Kakei 
and Michitomo Nishizawa for valuable discussions. 

K. F. was partially supported by Grant-in-Aid for Scientific Research (C) 
No. 10640210. \\
%%%%%%%%%%%%%%%%%%%%%%%%%%%%%%%%%%%%%%%%%%%%%%%%%%%%
%reference
  
\end{document}